\documentclass[aps,prl,twocolumn,groupedaddress,showpacs]{revtex4}
\usepackage{graphicx}
\usepackage{dcolumn}
\usepackage{amsmath}
\usepackage{amssymb}

\begin{document}

\title{Nonlinear viscoelastic dynamics of nano-confined water}

\author{Tai-De Li}
\affiliation{School of Physics, Georgia Institute of Technology,
Atlanta, GA 30332, USA}

\author{Elisa Riedo}
\affiliation{School of Physics, Georgia Institute of Technology,
Atlanta, GA 30332, USA}
\date{\today}

\vspace{1.5cm}
\begin{abstract}
The viscoelastic dynamics of nano-confined water is studied by
means of atomic force microscopy (AFM). We observe a nonlinear
viscoelastic behavior remarkably similar to that widely observed
in metastable complex fluids. We show that the origin of the
measured nonlinear viscoelasticity in nano-confined water is a
strain rate dependent relaxation time and slow dynamics. By
measuring the viscoelastic modulus at different frequencies and
strains, we find that the intrinsic relaxation time of
nano-confined water is in the range $0.1-0.0001$~s, orders of
magnitude longer than that of bulk water, and comparable to the
dielectric relaxation time measured in supercooled water at
$170-210$~K.
\end{abstract}

\maketitle

Confined fluids exhibit unique structural, dynamical,
electrokinetic, and mechanical properties that are different from
those of the
bulk~\cite{Tai-De07,ant01,ora04,uch05,mck05,zhu06,sch02}. Their
behavior depends on the degree of confinement, strain rate,
temperature, fluid molecular structure, and interactions with
boundaries. Surprising effects have been found when water is
confined in nanogaps. For example, the electric field induced
freezing of water at room temperature~\cite{cho05} and the
extremely high viscosity of water close to a mica
surface~\cite{Tai-De07,zhu01water,ant01}. Previous experiments and
calculations have pointed out the key role of the confining
surfaces~\cite{Tai-De07,uch05}. A notable increase in viscosity
and decrease in the diffusion constant was measured only when
water was confined between hydrophilic surfaces. For hydrophobic
confinement, the observed increase of viscosity was not very
pronounced. Intriguingly, a similar behavior has been observed in
confined glassy materials. When a glass-forming fluid is cooled
down to the glass transition temperature, $T_{g}$, its viscosity
grows by many orders of magnitude, and the confinement can
increase or decrease $T_{g}$ for strong or weak interactions with
the walls, respectively~\cite{sch02}.

So far, the viscosity measurements for nano-confined water have
been performed in the linear viscoelastic regime. However, as
observed in \textit{macroscopic} rheological measurements, the
study of the viscoelastic properties as a function of shear
amplitude and rate is important for a better understanding of the
dynamical and structural properties of fluids~\cite{Miyazaki06}.

In this letter, we investigate the viscoelastic response of
nano-confined water, as a function of shear amplitude and rate, by
means of direct high-resolution AFM measurements. We observe a
nonlinear viscoelastic behavior remarkably similar to that widely
observed in metastable complex fluids, such as gels and
supercooled liquids~\cite{Miyazaki06,nag96}. We show that the
origin of this nonlinear viscoelasticity in nano-confined water is
a strain rate dependent relaxation time and slow dynamics. By
measuring the viscoelastic modulus at different frequencies and
strains, we find that the intrinsic relaxation time, $\tau_{0}$,
of nano-confined water is in the range $0.1-0.0001$~s, orders of
magnitude longer than that of bulk water, and comparable to the
dielectric relaxation time measured in supercooled water at
$170-210$~K.~\cite{swen04}.

In our AFM experiments~\cite{szo05,Tai-De07}, a nano-size
spherical silicon tip is brought quasi-statically to the vicinity
of a flat freshly tape-cleaved hydrophilic mica surface, all
immersed in purified water, while small lateral oscillations are
applied to the cantilever support~\cite{Tai-De07}. The normal and
lateral forces acting on the tip are measured directly and
simultaneously as a function of the water film thickness, i.e.,
tip-sample distance, $d$. The zero distance, $d=0$, is where the
normal force diverges.

The experiments were performed with a Molecular Imaging PicoPlus
AFM. An instrumental problem in quasi-static force measurements is
that, during the tip-sample approach, the tip snaps into contact
with the surface at a distance where the gradient of the
tip-sample forces exceeds the cantilever normal spring constant,
$k_N$~\cite{rav01,osh92}. To overcome this problem, we employed
relatively stiff cantilevers. We used silicon tips with radii
$R=40\pm10$~nm and Ultrasharp NSC12/50 cantilevers with normal and
lateral spring constant in the range $k_N=3-4.5$~N/m and
$k_L=50-120$~N/m, respectively. Due to the mechanical stability of
our apparatus, and a judicious choice of the cantilever
stiffness~\cite{Tai-De07}, we are able to measure the tip-surface
distance with sub-Angstrom resolution all the way down to the last
adsorbed water layer. The calibration and force detection was
performed as described in \cite{szo05, lut95}. The approach
velocity was 0.2~nm/s. During the approach, lateral oscillations
parallel to the mica surface were applied to the cantilever holder
by means of a lock-in amplifier. The same lock-in amplifier was
then used to measure the amplitude of the lateral force, $F_{L}$,
and the phase difference, $\theta$, between the applied lateral
displacement and the detected lateral force. The zero of phase was
chosen when the tip was in hard contact with the mica surface, for
lateral oscillation amplitudes, $X_{0}$, small enough to guarantee
an elastic contact without slippage~\cite{car97}. In order to
shear parallel to the mica surface, before each measurement we
tilted the stage that holds the sample until the differences in
height of the mica surface topography across an area of $1\times
1$~$\mu$m$^{2}$ (as obtained from AFM sample topography imaging)
were smaller than $1$~nm. This corresponds to an angle $< 0.06
^{\rm o}$ between the mica surface and the tip shearing.

All the experiments were performed at 300~K in high purity DIUF
water from Fisher Chemicals (pH = 6.1). The purity of water used
in our AFM liquid cell was tested after the experiment by gas
chromatography - mass spectrometry (GC-MS). GC-MS spectra of used,
and not previously used, water samples were taken by 70SE
spectrometer (VG Instruments). In both cases the results showed
that any small molecular weight (less than 700 Da) organic
contaminants were present at amounts below the instrumental
threshold (5 ppm).

When a viscoelastic material is confined between two parallel
plates separated by $d$, with area $A$, and a sinusoidal strain is
applied to one of the plates at the frequency $\omega$,
$\gamma=\gamma_{0}\sin (\omega t)$, the resulting stress between
the plates can be written as $\sigma=\sigma _{0}\sin (\omega t +
\theta)$. The relationship between the strain amplitude,
$\gamma_0=\frac{X_0}{d}$, and the stress amplitude,  $\sigma
_0=\frac{F_L}{A}$, is given by the following equation:
\begin{equation}\label{eq:1}
\frac{F_L}{A}=\mid G^{*} \mid\frac{X_0}{d}
\end{equation}
where $G^{*}$ is the viscoelastic modulus. The viscoelastic
modulus contains information about the dissipative and elastic
response of the confined material. In particular, $G^{*}$ can be
written as a complex sum of the storage modulus, $G'$, and the
loss modulus, $G''$, i.e., $G^{*}=G'+iG''$, as~\cite{Ferry}:
\begin{equation}\label{eq:2}
G'=\frac{F_{L}d}{AX_{0}}\cos\theta,~
G''=\frac{F_{L}d}{AX_{0}}\sin\theta
\end{equation}
For a purely elastic solid, $\sigma$ and $\gamma$ remain in phase,
$\theta=0$, and so $G''=0$ and $G'=G^{*}$.

\begin{figure}
  \includegraphics[width=8cm]{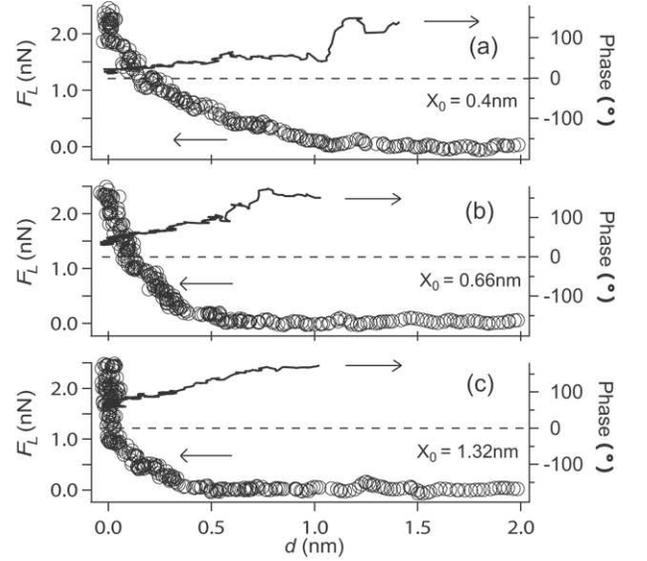}\\
  \caption{\small The
lateral force and phase as a function of tip-sample distance at
constant shear frequency, 955.3 Hz, and for three different shear
amplitudes, (a) $X_0=0.4$~nm. (b) $X_0=0.66$~nm. (c)
$X_0=1.32$~nm. The phase for $d>1$~nm is not shown because it
largely fluctuates in a random way.}\label{Fig. 1}
\end{figure}

In order to study the viscoelastic behavior of nano-confined water
we have measured the lateral force and the phase when we oscillate
laterally the AFM cantilever holder. As a first approximation we
have assumed that the lateral spring constant of our silicon
cantilever is much larger than the lateral tip-water contact
stiffness for $d<1$~nm~\cite{waterk}. As a consequence, we can
consider that the applied oscillation amplitude to the cantilever
holder is equal to the shear amplitude of the tip apex. Figure~1
shows $F_{L}$ and $\theta$ as a function of $d$ for three
different shear amplitudes at a fixed shear frequency,
$\omega=955.3$~Hz. For tip-sample distances larger than 1~nm we
observe that the lateral force is equal to zero within the
instrumental error for any shear amplitude. As soon as $d < $~1~nm
we observe that $F_{L}$ increases with decreasing $d$, and
diverges at $d = $~0~nm when the tip is in hard contact with the
mica surface. In a previous study~\cite{Tai-De07}, $F_{L}$ has
been used to calculate the viscosity of water ($\eta$) by using
Eq.~\ref{eq:1}, and by considering water as purely viscous, that
is, by making the approximation $\mid G^*  \mid\approx G'' \approx
\eta\cdot\omega$. This approximation is true when
$\theta\cong$~90$^o$, which, as we show later, is the case for
large strain rate amplitudes defined as
$\dot{\gamma_{0}}\equiv\gamma_{0}\cdot\omega$. However, the phase
measurements presented in Fig.~1 show that in general the behavior
of nano-confined water is viscoelastic, and furthermore, we
observe that the lateral force does not grow proportionally with
the shear amplitude, nor with the shear frequency (not shown
here). This is an indication that the viscoelastic response is not
linear, and the viscoelastic modulus is shear amplitude dependent,
$G^*=G^*(\gamma_{0})$. Therefore, a detailed study of $G^{*}$ as a
function of $\gamma_{0}$ is needed to shed light into this
nonlinear behavior.

\begin{figure}
  \includegraphics[width=8cm]{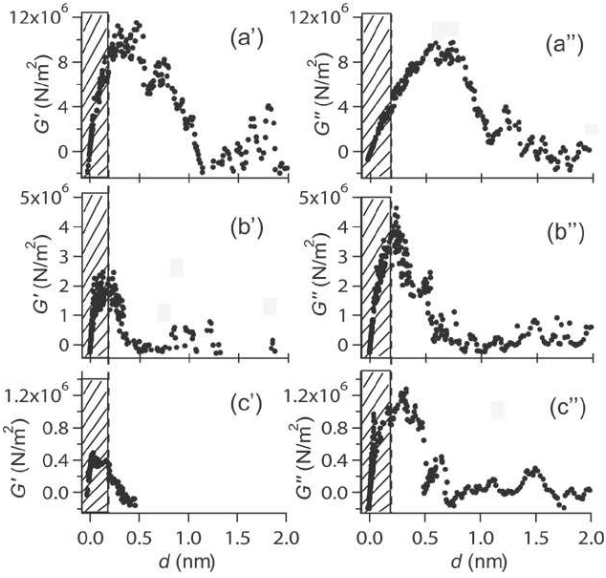}\\
  \caption{\small$G'$ and $G''$ as a function of tip-sample
distance. The shadowed area, $d<0.2$~nm, is not discussed in this
letter because the gap size is smaller than a water molecular
dimension. The frequency is $955.3$~Hz, and the shear amplitude is
$0.4$~nm for (a') and (a''), $0.66$~nm for (b') and (b''), and
$1.32$~nm for (c') and (c'').}\label{Fig. 2}
\end{figure}

By applying Eq.~\ref{eq:2} to the data in Fig. 1, we have
extracted $G'$ and $G''$ as a function of $d$ for different $X_0$
at a fixed $\omega$. (The $A$ used for Eq.~\ref{eq:2} is the
contact area corresponding to the spherical segment defined by the
intersection between the spherical tip and a plane at
$z=d+\triangle h$, $\triangle h=0.25$~nm, i.e., a water molecule
diameter~\cite{Tai-De07}.) Figure ~2 shows very clearly that $G'$
and $G''$ strongly depend on the shear amplitude. $G''$ dominates
over $G'$ for large shear amplitudes, where the response of
nano-confined water becomes purely viscous. Also, by decreasing
the gap size, we observe that the rise of $G'$ and $G''$ takes
place \textit{later} (smaller $d$) for larger shear amplitudes.
Furthermore, for all the investigated shear amplitudes, the rise
of $G''$ occurs \textit{earlier} (larger $d$) than the onset of
$G'$. The dramatic drop of both $G'$ and $G''$ for $d<0.2$~nm
(shadowed area in Fig.~2) is due to the invalidity of Eq.
\ref{eq:2} for distances smaller than the dimension of one water
molecule. Figure~2 indicates that the shear amplitude dependence
of the viscoelastic modulus is very complex and nonlinear. For
this reason we have performed measurements over a large range of
shear amplitudes and frequencies ($0.06$~nm$<X_{0}<2.8$~nm,
$50$~Hz~$<\omega<2$~kHz).

\begin{figure}
  \includegraphics[width=8cm]{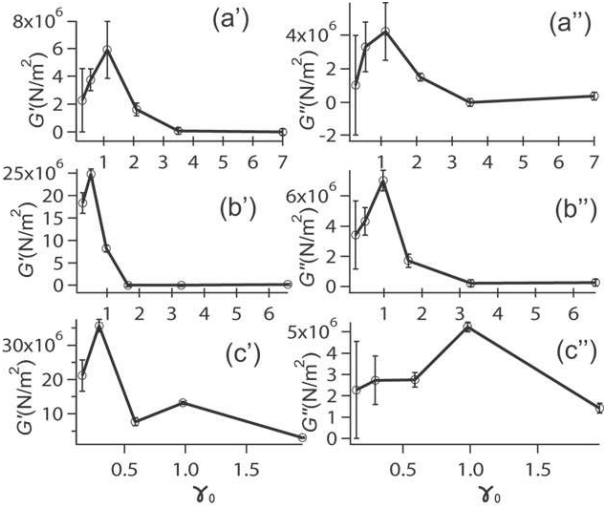}\\
  \caption{\small At
a constant distance, $d=0.4$~nm, $G'$ and $G''$ as a function of
$\gamma_0=\frac{X_{0}}{d}$. The shear frequency is $52.02$~Hz for
(a') and (a''), $955.3$~Hz for (b') and (b''), and $1.9689$~kHz
for (c') and (c'').}\label{Fig. 3}
\end{figure}

Following the Maxwell model for a linear viscoelastic system, the
relationship between the intrinsic relaxation time, $\tau_{0}$,
and the moduli, $G'$ and $G''$, is given by~\cite{Ferry}
\begin{equation}\label{eq:3}
G'=\frac{G_{0}(\omega\tau_{0})^{2}}{1+(\omega\tau_{0})^{2}},~
G''=\frac{G_{0}(\omega\tau_{0})}{1+(\omega\tau_{0})^{2}}
\end{equation}
where $G_{0}$ is a constant. According to Eq.~\ref{eq:3}, $G'$ and
$G''$ do not depend explicitly on $\gamma_0$. However, many
complex fluids experience a drastic decrease of their structural
relaxation time when they are subjected to large strains. This
phenomenon gives rise to a strong strain dependence of $G'$ and
$G''$, which can be described by the introduction of an
\textit{effective} relaxation time $\tau$ that depends on the
\emph{intrinsic} relaxation time and the strain
rate~\cite{Miyazaki06}. Once defined $\tau$, it is used to replace
$\tau_{0}$ in Eq.~\ref{eq:3}, and thus to predict $G'$ and $G''$
as a function of the strain. Recently, a phenomenological
expression has been found to characterize a strain dependent
effective relaxation time~\cite{Miyazaki06}
\begin{equation}\label{eq:4}
\frac{1}{\tau}\simeq\frac{1}{\tau_{0}}+K\cdot\dot{\gamma_0}^{\nu}
\end{equation}
where $\nu$ is a positive exponent, and $K$ is a constant. In a
glassy system which shows slow dynamics (
$\omega\gg\frac{1}{\tau_0}$), $\nu\sim1$ and
$K\sim1$~\cite{miy02,miy04}. By replacing $\tau_{0}$ in
Eq.~\ref{eq:3} with $\tau$ in Eq.~\ref{eq:4} when
$\omega\gg\frac{1}{\tau_0}$, the maximum of $G''$ is near
$\gamma_0\simeq1$, independently of the frequency. Figure~3
presents $G'$ and $G''$ vs. $\gamma_0$ for nano-confined water,
obtained by applying Eq.~\ref{eq:2} to the measured $F_L$ and
$\theta$ at three different frequencies for $d=0.4$~nm. $G'$ and
$G''$ in Fig.~3 show remarkable behavior: (i) the peak position of
$G''$ is around $\gamma_0\simeq1$ over a wide range of
frequencies; (ii) for $\gamma_0<1$, the viscoelasticity is
dominantly elastic, i.e., $G'>G''$; and (iii) $G'$ and $G''$ decay
to zero for large values of $\gamma_0$. These features of our
nano-confined water system are ubiquitous in metastable complex
fluids~\cite{Miyazaki06} and they are captured by the argument of
the strain rate dependent effective relaxation time. Indeed, by
using Eq.~\ref{eq:3} and~\ref{eq:4} the shape of the curves
presented in Fig.~3 can be fully described.

\begin{figure}
  \includegraphics[width=8cm]{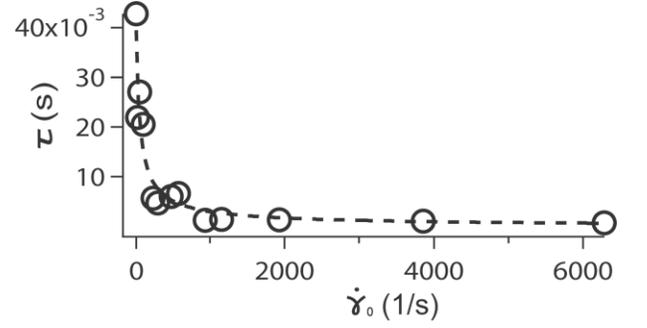}\\
  \caption{\small
The effective relaxation time as a function of
$\dot{\gamma_0}=\gamma_0\cdot\omega$ at $d=0.4$~nm. The dash line
is the fitting with Eq.~\ref{eq:4}.}\label{Fig. 4}
\end{figure}

From Eq.~\ref{eq:3}, the effective relaxation time can be
predicted by:
\begin{equation}\label{eq:5}
\tau=\frac{G'}{G''}\cdot\frac{1}{\omega}
\end{equation}
By using Eq.~\ref{eq:5} and the experimental values of $G'$ and
$G''$, $\tau$ as a function of $\dot{\gamma_0}$ for $d=0.4$~nm is
determined and shown in Fig.~4. The effective relaxation time of
nano-confined water decreases from $40$~ms to $0.7$~ms when the
strain rate increases from $14$~s$^{-1}$ to $6000$~s$^{-1}$. The
nonlinearity of the relaxation time sets in when the experimental
time scale ($\dot{\gamma_0}$) is faster than the intrinsic
relaxation time ($\tau_0$). In this case, the time response can
only be measured effectively as a function of the experimental
time scale.

By fitting the data in Fig.~4 with Eq.~\ref{eq:4} we found that
the intrinsic relaxation time was $\tau_0=0.06\pm0.03$~s for
$K=0.95\pm1.49$ and $\nu=0.84\pm0.29$. These values of $\nu$ and
$K$ are similar to those found in complex metastable fluids with
slow dynamics~\cite{miy02,miy04}. The striking result is that the
observed $\tau$ and $\tau _{0}$ are orders of magnitude slower
than the relaxation time of bulk water at room temperature, which
is of the order of $10^{-12}$~s. The fact that confinement can
drastically slow down the dynamics of a fluid has been previously
observed in diverse systems, such as colloidal
suspensions~\cite{weeks07} and polymers~\cite{mck05}, where for
strong fluid-wall interactions, the glass transition temperature
is shifted towards high temperatures upon
confinement~\cite{sch02}. An alternative way to view this behavior
is to consider that the confinement defines an effective
temperature of the system which is lower than the canonical
temperature. According to a previous study~\cite{swen04}, the
dielectric relaxation time of supercooled water confined in clays
at $175$~K is about $0.06$~s similar to the relaxation time found
in our experiments on nano-confined water at room temperature.
Moreover, the value of the viscosity measured in our
investigations is comparable with that of supercooled water at
$140$~K in a $100$~$\mu$m radius tube~\cite{hal63}. A recent study
has shown that the dielectric relaxation time of supercooled water
is very sensitive to the confinement~\cite{swen06}. For
confinement lengths of the order of 1 nm, it was found that, over
a wide range of temperatures, the dielectric relaxation times are
always longer than in bulk water. In our experiments, we also
observe that $\tau$ decreases with reduced confinement, i.e., with
increasing $d$. Unfortunately, for $d\geq1$~nm the lateral force
becomes so small that we cannot measure its value precisely due to
low signal-to-noise ratio. The only information that we can
extract is that the intrinsic relaxation time for $d\geq1$~nm is
shorter than $10^{-4}$~s.

In conclusion, we have studied the room temperature viscoelastic
properties of nano-confined water, finding a slow dynamical
behavior similar to that observed in metastable complex fluids. By
measuring the viscoelastic modulus at different frequencies and
strains, we have found that the intrinsic relaxation time of
nano-confined water is about $0.06$~s. This value is comparable
with the dielectric relaxation time measured in supercooled water
at $175$~K.

We thank H. M. Wyss and L. Bocquet for helpful discussions. We
gratefully acknowledge the financial support of the NSF
DMR-0405319, DOE DE-FG02-06ER46293 and NSF STC grants.


\begin{thebibliography}{24}
\bibitem[{wat()}]{waterk}
\bibinfo{note}{From the results in Fig.~2, the shear elastic mudulus of
  nanoconfined water, $G'$, is three orders of magnutide smaller than the shear
  modulus of silicon, $G^s\simeq50$~GPa. The tip-water contact lateral
  stiffness is approximately equal to
  $K_{L}^{c}\sim8a(\frac{1}{G'}+\frac{1}{G^{s}})^{-1}\sim0.4$~N/m, ($a$ is the
  contact radius, $a\simeq4.5$~nm in these experiments), which is two orders of
  magnitude smaller than the torsion and lateral force constant.}
\end{thebibliography}
\end{document}